\documentclass[12pt]{article}
\usepackage{amssymb,amsmath,epsfig}
\usepackage{cite}
\usepackage[utf8]{inputenc}
 \allowdisplaybreaks
\begin{document}
\title{\bf Compact relativistic geometries in $f(R,G)$ gravity}
\author{W. U. Rahman$^1$ \thanks{Waheed71rahman@gmail.com},
M. Ilyas$^2$ \thanks{ilyas\_mia@yahoo.com},
Z. Yousaf$^3$\thanks{zeeshan.math@pu.edu.pk},
S. Ullah$^2$\thanks{saeedullah.phy@gmail.com},
F. Khan$^2$\thanks{fawad.ccms@gmail.com} and
R. khan$^1$\thanks{rashidkhan@awkum.edu.pk}\\
$^1$ Department of Physics, Abdul Wali khan University Mardan,\\
Mardan, 23200, Pakistan.\\
$^2$ Institute of Physics, Gomal University,\\
Dera Ismail Khan, 29220,
Khyber Pakhtunkhwa, Pakistan.\\
$^3$ Department of Mathematics, University of the Punjab,\\
Quaid-i-Azam Campus, Lahore-54590, Pakistan.}
\date{}
\maketitle

\begin{abstract}
One of the possible potential candidates for describing the universe's rapid expansion is modified gravity. In the framework of the modified theory of gravity $f(R,G)$, the present work features the materialization of anisotropic matter, such as compact stars. Specifically, to learn more about the physical behavior of compact stars, the radial, and tangential pressures as well as the energy density of six stars namely $Her X-1$, $SAXJ1808.4-3658$, $4U1820-30$, $PSR J 1614 2230$, $VELA X-1$, and $Cen X-3$ are calculated. Herein, the modified theory of gravity $f(R,G)$ is disintegrated into two parts i.e. the $\tanh$ hyperbolic $f(R)$ model and the three different $f(G)$ model. The study focuses on graphical analysis of compact stars wherein the stability aspects, energy conditions, and anisotropic measurements are mainly addressed. Our calculation revealed that, for the positive value of parameter n of the model $f(G)$, all the six stars behave normally.
\end{abstract}
\section{Introduction}
The spatial behavior of the components of the universe is listed among one of the complex phenomena. The observation of the universe not only educates us about its continuous expansion but also motivates the scientific community to find alternate justification which may be more beneficial to enlighten the phenomenon of the current cosmic expansion \cite{ref1}. Extensive theoretical studies, focusing on the expansion of the universe, has been carried out where several alternative models currently known as the modified theories of general relativity were developed. These modified theories of gravity include $f(R)$, $f(R,T)$,$ f(G)$, $f(G,T)$, $f(T)$, $f(R,G)$ and $f(R,G,T)$ where $R$ is the Ricci-scalar term, $T$ represents the trace of the energy-momentum tensor, while the Gauss-Bonnet term is denoted by $G$. Some alterations are highly desirable to explain the scenario of universe expansion in the strong-field regime as the theory of general relativity mainly focuses on the cosmological phenomena in the weak-field regime. In 1970, Buchdahl was motivated by the same thought and he proposed one of the modified theories i.e. the $f(R)$ gravity \cite{ref2} which is the most basic modification of general relativity and has been widely studied in terms of neutron and compact stars stability and existence. The Lane-Emden equation was tackled to explore the stellar structure and hydrostatic equilibrium in the $f(R)$ gravity \cite{ref3}, for further review, see \cite{ref3a,ref3b}. The higher-derivative gravitational invariants were used for the description of the finite-time future singularities along with their solution in the modified gravity \cite{ref4,ref3a}. In 2011, Harko et al. adduced the matter and curvature terms to present the modified theory $f(R,T)$ \cite{ref5} where the dependency of $T$ may be caused by the quantum effects or exotic imperfect fluids. For works discussing the related possibility of direct detection of dark energy, see \cite{ref5a}. Moreover, a further amendment to general relativity is, the Gauss-Bonnet gravity \cite{ref6}, which is one of the notorious family of modified theories and is also termed the Einstein-Gauss-Bonnet gravity. In this theory, the Hilbert-Einstein action was modified to include the Gauss-Bonnet term
\begin{equation}
G = R^2 + R_{\mu\nu\theta\phi} R^{\mu\nu\theta\phi}-4R_{\mu\nu}R^{\mu\nu}
\end{equation}
where $R_{\mu\nu\theta\phi}$ and $R_{\mu\nu}$ are the Riemann and Ricci tensors respectively. The extra Gauss-Bonnet term was used to resolve the deficiencies in the $f(R)$ theory of gravity and has been the focus of intense research \cite{A2,A3,A3a,A3b,A3c,A3d}. The generalized form of Gauss-Bonnet gravity is the $f(G)$ gravity which is capable of reproducing all types of cosmological solutions \cite{ref7,ref7a}. The $f(G)$ model, being less constrained as compared to $f(R)$ model, is more advantageous \cite{A4}. Additionally, as an alternative to dark energy, the $f(G)$ model provides an effective platform for analyzing a variety of cosmic issues \cite{A5}.
Recently, Sharif et al. suggested an alternative theory, known as $f(G,T)$ gravity, and investigated the energy conditions for the eminent Friedmann-Robertson-Walker space-time. This theory has drawn considerable attention and has become the focus of groundbreaking work \cite{ref10,ref11,ref12,ref13}. Torsion scalar, rather than the curvature, describes the gravitational interaction in another modified gravity theory known as teleparallel gravity (which has gained popularity in recent years) \cite{A6,A7}. Motivated by the generalization of $f(R)$ gravity, the teleparallel gravity was generalized by replacing the torsion-scalar with an analytic function of the torsion-scalar \cite{ref14}. This modified theory of gravity, known as $f(T)$ gravity, is more advantageous over the $f(R)$ gravity because the field-equations in the latter turn out to be differential equations of fourth-order while the field-equations in the former are second-order differential equations which are easy to handle \cite{A8}.
An attempt was made to combine the $R$ and $G$ in a bivariate function $f(R,G)$ \cite{ref15,ref16,ref17,ref18,ref19,ref20,ref21,ref22,ref22a} which provides a basis for the double inflationary scenario \cite{ref23} and is strongly supported by the observational finding \cite{ref24,ref25}. In addition to its stability, the $f(R,G)$ theory is well-suited to describe the crossing of the phantom divide line, as well as the accelerated waves of celestial bodies and transformation from the accelerating to decelerating phases. The reduction in risk of the ghost contributions is one of the essential features of the $f(R,G)$ theory of gravity where the term $G$ was introduced to regularize the gravitational action \cite{ref26,ref27}. Hence, the modified theories of gravity appear to be appealing for describing the universe in various cosmological contexts \cite{ref28}.
Compact stars have always been the focus of intensive research \cite{ref29,ref30,ref31,ref32,ref33,ref34,ref35,ref36,ref37,ref38,ref39,ref40,ref40a,ref40b,ref40c,ref40d,ref40e,ref40f,ref40g,ref40h}. Various properties of neutron stars, such as radius, mass, and the moment of inertia, have been investigated, and comparisons with the theory of general relativity and alternative gravity theories have been established \cite{ref41}. Two distinct hadronic parameters and the strange-matter equation-of-state parameter have been used to investigate the structure of slowly spinning neutron stars in the $R^2$-gravity \cite{ref42,ref42a,ref42b,ref42c,ref42d}. For inflationary models in the context of scalar tensor gravity with $R^2$-gravity type, see \cite{ref42e,ref42f,ref42g}. Furthermore, in the presence of cosmological constants, the mass ratio of compact bodies has been calculated in Ref. \cite{ref43}. The present work aims to study the relativistic geometries by considering different viable models in the $f(R,G)$ gravity.\\

The following is the format of this paper:\\
In section 2, we present the basic formulism of $f(R,G)$ gravity, its equation of motions, and the analytic solution for the anisotropic matter distribution under different viable models in $f (R,G)$ gravity. Section 3 deals with the physical analysis of the given system which consist of six different relativistic geometries, namely Her Stars $Her X-1$, $SAXJ1808.4-3658$, $4U1820-30$, $PSR J 1614 2230$, $VELA X-1$ and $Cen X-3$. The last section summarizes the results of the paper.

\section{The Modified $f(R,G)$ gravitational theory}
We shal start with the operation of Gauss-Bonnet's gravity. \cite{ref44}.
\begin{equation}
I = \int {d{x^4}\sqrt { - g} \left[ {f(R,G) + {L_m}} \right]}
\end{equation}
The following modified equations in the field are obtained by varying the operation $(2)$ with respect to the yields of the metric-tensor. \cite{ref25}.
\begin{equation}
\begin{gathered}
{R_{\mu \nu }} - \frac{1}{2}{g_{\mu \nu }}R = {T_{\mu \nu }} + {\nabla _\mu }{\nabla _\nu }{f_R} - {g_{\mu \nu }}\Box{f_R} + 2R{\nabla _\mu }{\nabla _\nu }{f_G} - 2{g_{\mu \nu }}R \Box{f_G}\\
- 4R_\mu ^\alpha {\nabla _\alpha }{\nabla _\nu }{f_G} - 4R_\nu ^\alpha {\nabla _\alpha }{\nabla _\mu }{f_G} + 4{R_{\mu \nu }}\Box{f_G} + 4{g_{\mu \nu }}{R^{\alpha \beta }}{\nabla _{^\alpha }}{\nabla _\beta }{f_G}\\
+ 4{R_{\mu \alpha \beta \nu }}{\nabla ^\alpha }{\nabla ^\beta }{f_G} - \frac{1}{2}g(R{f_R} + G{f_G} - f(R,G))
\end{gathered}
\end{equation}
Where partial derivatives are $f_{R}$ and $f_{G}$ with respect to $R$ and $G$, respectively,
\begin{equation}
{T_{\mu \nu }} = (\rho + p){U_\mu }{U_\nu } - {p_t}{g_{\mu \nu }} + ({p_r} - {p_r}){V_\mu }{V_\nu }
\end{equation}
in which $U_\mu$ is the fluid of four velocity and $U_\mu U^\mu=1$ and $V_\mu V^\mu=-1$.
\subsection{Matter distribution}
The general form for line element for dynamical spherical symmetric geometry is.
\begin{equation}
d{s^2} = e^{a}d{t^2} -e^{b}d{r^2}-{r^2}( {d{\theta ^2} + {{r^2}{\sin }^2}\theta d{\phi ^2}}),
\end{equation}
where $a$ and $b$ are the functions of radial coordinate and we assume $a(r) =r^2 B + C$ and $b(r)= r^2 A$ (Krori-Barua (1975) solution), $A$, $B$ and $C$ are arbitrary constants that can be determined using physical conditions.
Now solving the field equations, we get
\begin{equation}
\begin{gathered}
\rho = \frac{1}{{2{r^2}}}{{\rm{e}}^{ - 2b}}( - {{\rm{e}}^{2b}}{r^2}f(R,G) + {{\rm{e}}^{2b}}{r^2}G{f_G} + {{\rm{e}}^b}{f_R}( - 2 + 2{{\rm{e}}^b} + {{\rm{e}}^b}{r^2}R + 2rb')\\
+ 12b'{f_G}^\prime - 4{{\rm{e}}^b}b'{f_G}^\prime - 4{{\rm{e}}^b}r{f_R}^\prime + {{\rm{e}}^b}{r^2}b'{f_R}^\prime - 8{f_G}^{\prime \prime } + 8{{\rm{e}}^b}{f_G}^{\prime \prime } - 2{{\rm{e}}^b}{r^2}{f_R}^{\prime \prime })
\end{gathered}
\end{equation}
\begin{equation}
\begin{gathered}
p_r = \frac{1}{{2{r^2}}}{{\rm{e}}^{ - 2b}}({{\rm{e}}^{2b}}{r^2}f(R,G) - {{\rm{e}}^{2b}}{r^2}G{f_G} - {{\rm{e}}^b}{f_R}( - 2 + 2{{\rm{e}}^b}\\
+ {{\rm{e}}^b}{r^2}R - 2ra') + 12a'{f_G}^\prime - 4{{\rm{e}}^b}a'{f_G}^\prime + 4{{\rm{e}}^b}r{f_R}^\prime + {{\rm{e}}^b}{r^2}a'{f_R}^\prime )
\end{gathered}
\end{equation}
and
\begin{equation}
\begin{gathered}
p_t = \frac{1}{{4r}}{{\rm{e}}^{ - 2b}}(2{{\rm{e}}^{2b}}rf(R,G) - 2{{\rm{e}}^{2b}}rG{f_G} - 2{{\rm{e}}^{2b}}r{f_R}R + 2{{\rm{e}}^b}{f_R}a'\\
+ {{\rm{e}}^b}r{f_R}{{a'}^2} - 2{{\rm{e}}^b}{f_R}b' - {{\rm{e}}^b}r{f_R}a'b' + 4{{a'}^2}{f_G}^\prime - 12a'b'{f_G}^\prime + 4{{\rm{e}}^b}{f_R}^\prime \\
+ 2{{\rm{e}}^b}ra'{f_R}^\prime - 2{{\rm{e}}^b}rb'{f_R}^\prime + 2{{\rm{e}}^b}r{f_R}{a^{\prime \prime }} + 8{f_G}^\prime {a^{\prime \prime }} + 8a'{f_G}^{\prime \prime } + 4{{\rm{e}}^b}r{f_R}^{\prime \prime })
\end{gathered}
\end{equation}
where the ricci scalar $R$ is
\begin{equation}
R = \frac{{{{\rm{e}}^{ - b}}}}{{2{r^2}}}(4 - 4{{\rm{e}}^b} + {r^2}{a'^2} - 4rb' + ra'\left( {4 - rb'} \right) + 2{r^2}{a^{\prime \prime }})
\end{equation}
and
\begin{equation}
G = \frac{{2{{\rm{e}}^{ - 2b}}}}{{{r^2}}}\left( { - \left( { - 1 + {{\rm{e}}^b}} \right){{a'}^2} + \left( { - 3 + {{\rm{e}}^b}} \right)a'b' - 2\left( { - 1 + {{\rm{e}}^b}} \right){a^{\prime \prime }}} \right)
\end{equation}
using the definition of $a$ and $b$
\begin{equation}\label{ro}\nonumber
\begin{split}
\rho = \frac{1}{{2{r^2}}}{{\rm{e}}^{ - 2A{r^2}}}({{\rm{e}}^{2A{r^2}}}(2{f_R} + {r^2}( - f(R,G) + G{f_G} + {f_R}R)) + 24Ar{f_G}^\prime - 8{f_G}^{\prime \prime}\\
+ 2{{\rm{e}}^{A{r^2}}}(\left( { - 1 + 2A{r^2}} \right){f_R} + 4{f_G}^{\prime \prime } + r( - 4A{f_G}^\prime + \left( { - 2 + A{r^2}} \right){f_R}^\prime - r{f_R}^{\prime \prime })))
\end{split}
\end{equation}
\begin{equation}\label{pr}
\begin{gathered}
{p_r} = \frac{1}{{2{r^2}}}{{\rm{e}}^{ - 2A{r^2}}}({{\rm{e}}^{2A{r^2}}}( - 2{f_R} + {r^2}(f(R,G) - {f_G}G - {f_R}R)) + 24Br{f_G}^\prime \\
+ 2{{\rm{e}}^{A{r^2}}}\left( {{f_R} + 2B{r^2}{f_R} - 4Br{f_G}^\prime + r\left( {2 + B{r^2}} \right){f_R}^\prime } \right))
\end{gathered}
\end{equation}
\begin{equation}\label{pt}
\begin{gathered}
{p_t} = \frac{1}{{2r}}{{\rm{e}}^{ - 2A{r^2}}}({{\rm{e}}^{2A{r^2}}}r(f(R,G) - {f_G}G - {f_R}R) + 8B\left( {\left( {1 + \left( { - 3A + B} \right){r^2}} \right){f_G}^\prime + r{f_G}^{\prime \prime }} \right)\\
- 2{{\rm{e}}^{A{r^2}}}\left( {r\left( {A - 2B + \left( {A - B} \right)B{r^2}} \right)fR + \left( { - 1 + \left( {A - B} \right){r^2}} \right){f_R}^\prime - r{f_R}^{\prime \prime }} \right))
\end{gathered}
\end{equation}
With the help of these equations, we developed the theoretical modeling of compact geometries, where the values of constants $A,B,$ and $C$ will be found in the following section.
\subsection{Interior-Exterior boundary conditions}
The interior metric is compared with the vacuum solution of exterior spherical symmetric metric, found as
\begin{equation}
d{s^2} = \left[ {1 - \frac{{2M}}{r}} \right]d{t^2} - {\left[ {1 - \frac{{2M}}{r}} \right]^{ - 1}}d{r^2} - {r^2}\left( {d{\theta ^2} + {\mathop{\rm \sin}\nolimits} {\theta ^2}d{\varphi ^2}} \right)
\end{equation}
\begin{equation}
A = \frac{{ - 1}}{{{R^2}}}\ln \left[ {1 - \frac{{2M}}{r}} \right],
\end{equation}
\begin{equation}
B = \frac{M}{{{R^3}}}{\left[ {1 - \frac{{2M}}{r}} \right]^{ - 1}},
\end{equation}
\begin{equation}
C = \ln \left[ {1 - \frac{{2M}}{r}} \right] - \frac{M}{R}{\left[ {1 - \frac{{2M}}{r}} \right]^{ - 1}}.
\end{equation}
With the help of these equations, we assumed six different compact stars and found the numerical values of $A$ and $B$, as listed in Table: \ref{table:1} below
\begin{table}[h!]
\centering
\begin{tabular}{||c|| c|| c|| c|| c|| c||}
\hline
Compact Stars $(CSi)$&M $(M_{\odot})$ & R $(km)$ & $\mu=\frac{M}{R}$ &A $(km^{-2})$ &B $(km^{-2})$ \\ [0.5ex]
\hline\hline\
$Her X-1$ (CS1) & $0.88$ & $7.7$ & $0.168$ & $0.006906276428 $ &$0.004267364618 $\\ [1ex]
\hline\
$SAXJ1808.4-3658$ (CS2)& $1.435$ & $7.07$ & $0.299$ & $0.01823156974$ &$0.01488011569$\\ [1ex]
\hline
$4U1820-30$ (CS3)& $2,25$ & $10$ & $0.332$ & $0.01090644119 $ &$0.009880952381$\\ [1ex]
\hline\
$PSR J 1614 2230$ (CS4)& $1.97$ & $10.977$ & $0.1795$ & $0.003689961987 $
&$0.002323332389 $\\ [1ex]
\hline\
$VELA X-1$ (CS5)& $1.77$& $10.654$ & $0.1661$ & $0.003558090580 $ &$0.002191967045 $\\ [1ex]
\hline
$Cen X-3$ (CS6)& $1.49$ & $10.136$ & $0.1471$ & $0.003388625404 $
&$0.002026668572 $\\ [1ex] \hline
\end{tabular}
\caption{The estimated quantities of masses $M$, radii $R$, compactness $\mu$, and constants $A$, and $B$ of the six compact stars}
\label{table:1}
\end{table}
\subsection{Different Models}
For the sake of convenience, in this section, we used some of the more realistic models for studying various properties of compact star such as mass, energy conditions etc.
$$f(R,G)=f_1(R)+f_2(G)$$
where we take $f_1(R)=R + \alpha \delta R \tanh(R/\delta)$ \cite{ref44a}
\subsubsection{Model 1}
\begin{equation}\label{model1}
f(R,G) = R + \alpha \delta R \tanh(R/\delta) + \beta {G^\lambda } + \gamma G\log \left[ G \right]
\end{equation}
Where $\alpha, \delta, \beta$ and $\lambda$ are the parameters of the desire model. The second term is taken from Ref. \cite{ref44b}.
\subsubsection{Model 2}
We have used the following viable model as
\begin{equation}\label{model2}
f(R,G) =R + \alpha \delta R \tanh(R/\delta) + \beta {G^n}\left( {1 + \gamma {G^m}} \right)
\end{equation}
Where $\alpha, \delta, \beta, \gamma, m,$ and $n$ are the desire model parameters. The second term is taken Ref. \cite{ref44c}.
\subsubsection{Model 3}
we have considered the following viable model
\begin{equation}\label{model3}
f(R,G) = R + \alpha \delta R \tanh(R/\delta) + \frac{{{a_1}{G^n} + {b_1})}}{{\left( {{a_2}{G^n} + {b_2}} \right)}}
\end{equation}
Where $\alpha, \delta, \beta, a_1, a_2, b_1, b_2$, and $n$ are the parameters of the desire model. The second term is taken from Ref. \cite{ref44d}.
\section{Physical Aspects of the Models}
Here, we have gone through some of the physical requirements for the interior approach. The anisotropic activity and stability parameters are presented in the following sections.
\subsection{Energy Density and Pressure Evolutions}
\begin{figure}[h!]
\centering
\includegraphics[width=20pc]{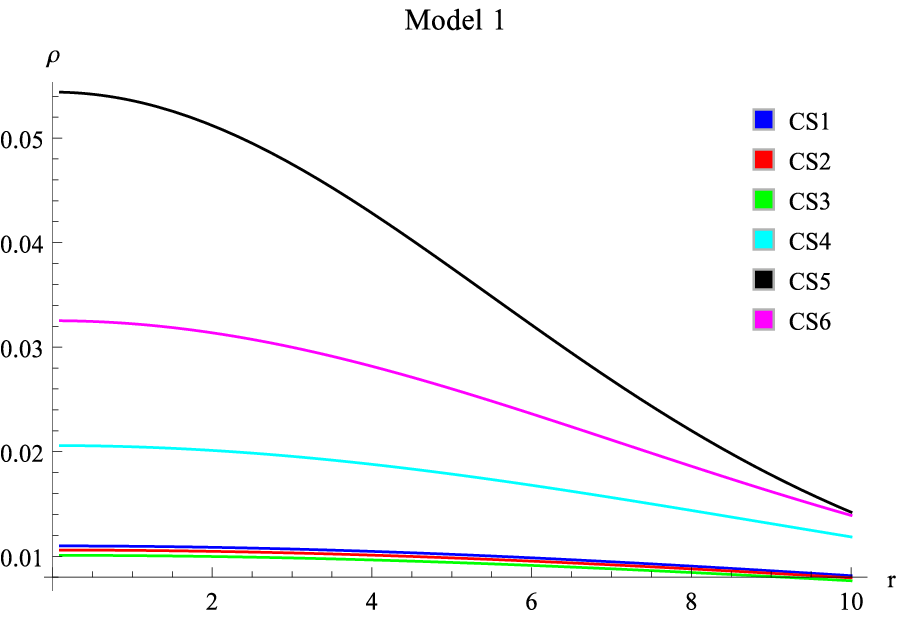}
\includegraphics[width=20pc]{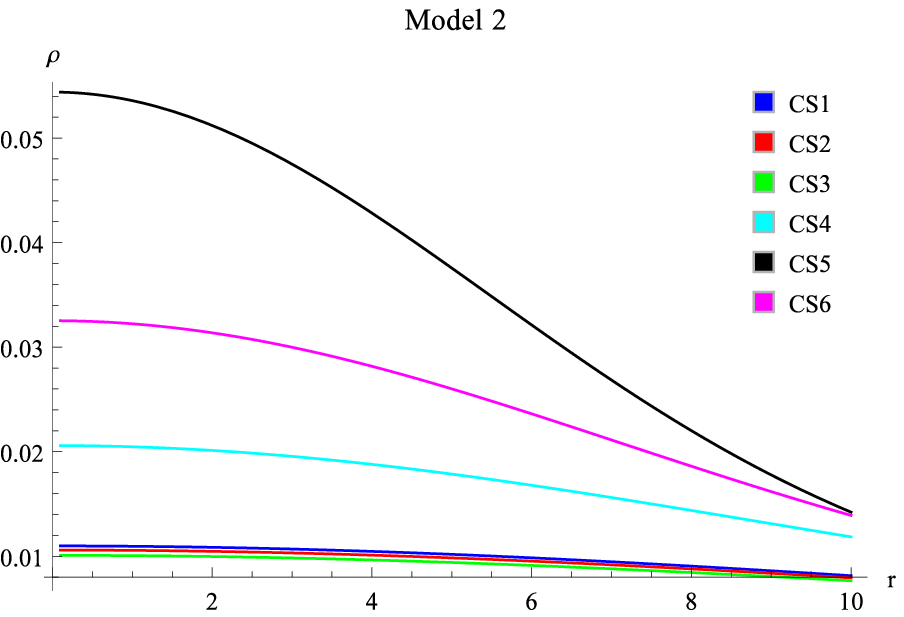}
\includegraphics[width=20pc]{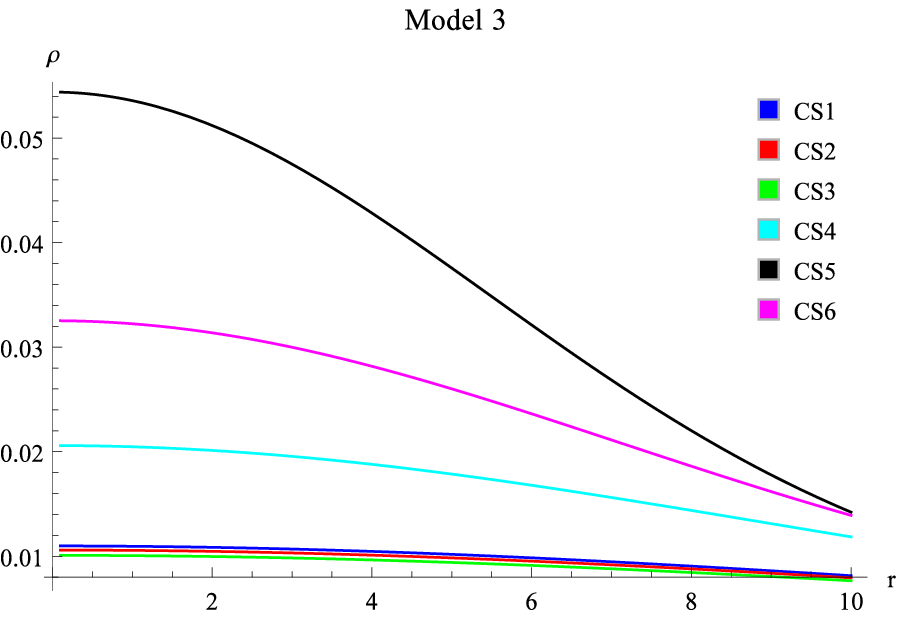}
\caption{The density $(km^{-2})$ variation of strange compact stars candidate, $Her X-1$, $SAXJ1808.4-3658$, $4U1820-30$, $PSR J 1614 2230$, $VELA X-1$ and $Cen X-3$ under three different viable Models.}\label{roc}
\end{figure}

\begin{figure}[h!]
\centering
\includegraphics[width=20pc]{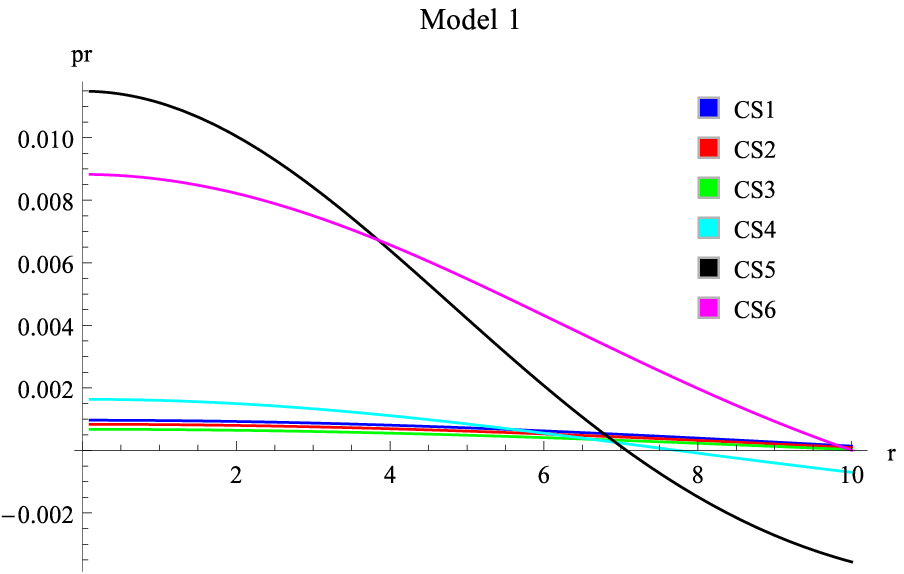}
\includegraphics[width=20pc]{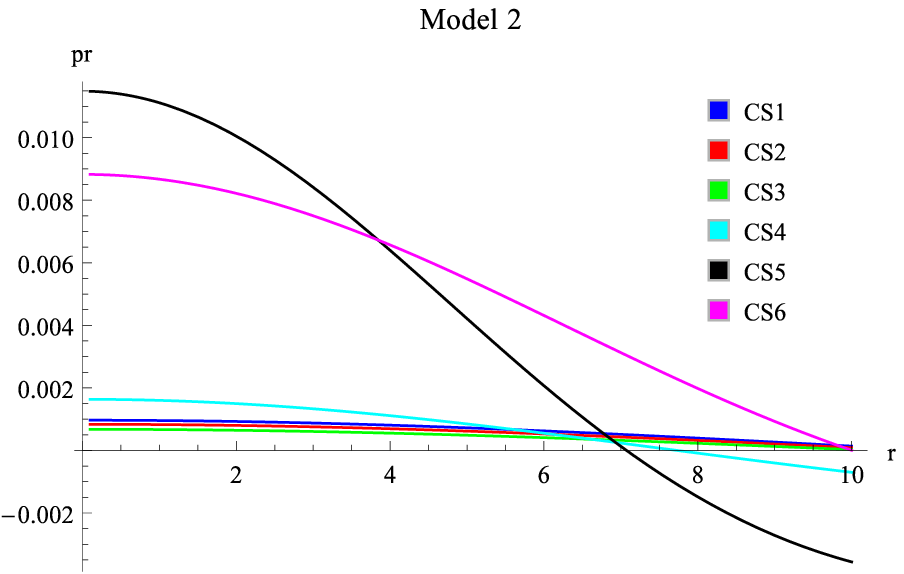}
\includegraphics[width=20pc]{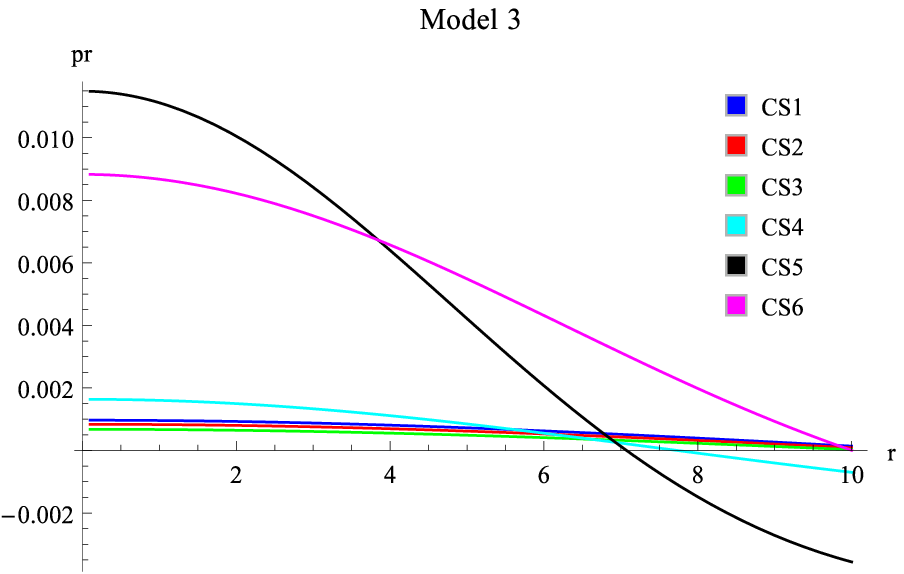}
\caption{The radial pressure $(km^{-2})$ variation of strange compact stars candidate, $Her X-1$, $SAXJ1808.4-3658$, $4U1820-30$, $PSR J 1614 2230$, $VELA X-1$ and $Cen X-3$ under three different viable Models.}\label{prc}
\end{figure}

\begin{figure}[h!]
\centering
\includegraphics[width=20pc]{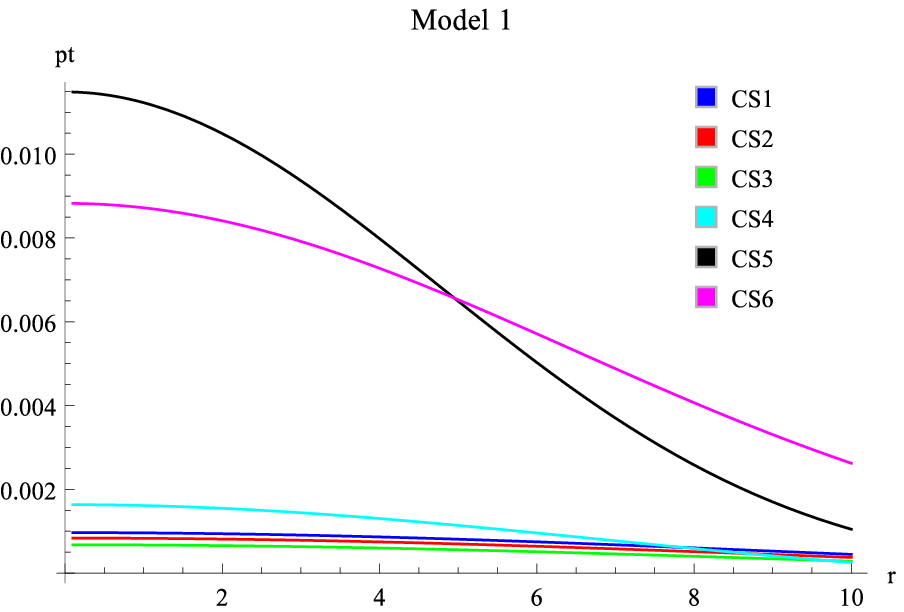}
\includegraphics[width=20pc]{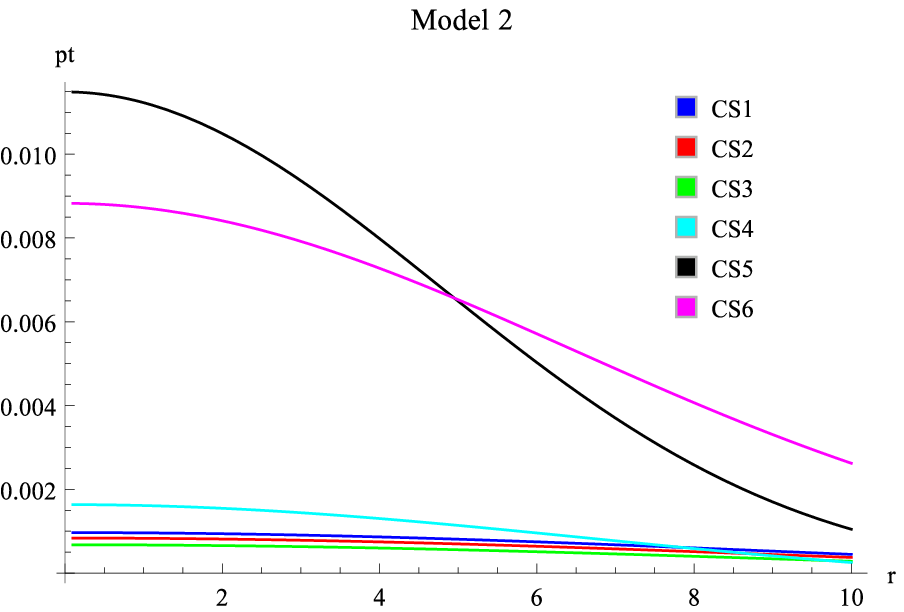}
\includegraphics[width=20pc]{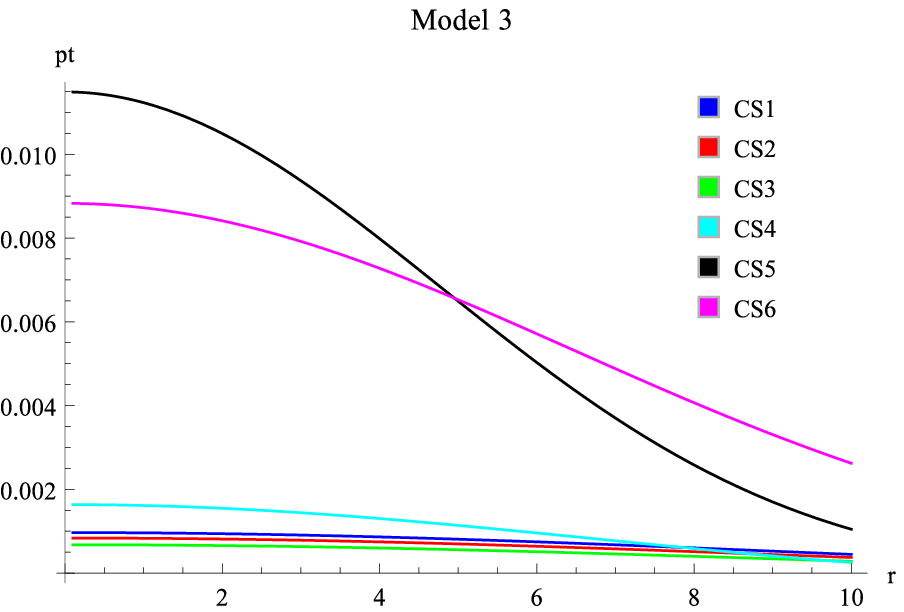}
\caption{The transverse pressure $(km^{-2})$ variation of strange compact stars candidate, $Her X-1$, $SAXJ1808.4-3658$, $4U1820-30$, $PSR J 1614 2230$, $VELA X-1$ and $Cen X-3$ under three different viable Models.}\label{ptc}
\end{figure}

The action of energy density, anisotropic stresses, and their radial derivative for each compact star was obtained by using the numerical values of constant $A$ and $B$ into the field equations along with the suggested three related models. These variations can be observed in Fig. (\ref{roc}-\ref{ptc}). This shows that as $r\rightarrow0$ the density $\rho$ goes to maximum. Simply the $\rho$ is like the decreasing function of $r$ e.g. $\rho$ is decreasing with the increase of $r$ which, in fact, reflects the highly compact star's core, indicating that our $f(R,G)$ models are viable for the core's outer area.\\
The central density of each star is maximum while minimum at the surface of stars e.g. $1.10347\times10^{15} g/cm^3$ is the central density of star $Her X-1$ while its surface density is $7.88953\times10^{14} g/cm^3 $.\\
Similarly, the compact star, $SAXJ1808.4-3658$ has $2.91534\times10^{15} g/cm^3$ central density and its surface density is $1.41936\times10^{15} g/cm^3 $.\\
Furthermore, the compact star $4U1820-30$ has a central density of $1.74345\times10^{15}g/cm^3$ and the surface density of $7.448578\times10^{14}g/cm^3$.\\
The surface and central densities are summarized in Table \ref{table:2}.
\begin{table}[h!]
\footnotesize
\tiny
\setlength{\tabcolsep}{-0em}
\begin{tabular}{|c|| c |c |c |c |c| c||}
\hline
Compact Stars & Central density Model 1& Surface density Model 1& Central density Model 2& Surface density Model 2& Central density Model 3& Surface density Model 3\\
\hline
Cs 1& $1.10347 \times 10^{15}$& $7.88953 \times 10^{14}$& $1.10343 \times 10^{15}$& $7.88928 \times 10^{14}$& $1.10343 \times 10^{15}$& $7.88928 \times 10^{14}$\\ [1ex]
\hline
Cs 2& $2.91534 \times 10^{15}$& $1.41936 \times 10^{15}$& $2.91534 \times 10^{15}$& $1.41935 \times 10^{15}$& $2.9154 \times 10^{15}$& $1.41936 \times 10^{15}$\\ [1ex]
\hline
Cs 3& $1.7435 \times 10^{15}$& $7.44858 \times 10^{14}$& $1.74341 \times 10^{15}$& $7.44863 \times 10^{14}$& $1.74343 \times 10^{15}$& $7.44882 \times 10^{14}$\\ [1ex]
\hline
Cs 4& $5.88904 \times 10^{14}$& $4.08874 \times 10^{14}$& $5.88861 \times 10^{14}$& $4.08891 \times 10^{14}$& $5.88845 \times 10^{14}$& $4.0893 \times 10^{14}$\\ [1ex]
\hline
Cs 5& $5.67763 \times 10^{14}$& $4.07021 \times 10^{14}$& $5.67763 \times 10^{14}$& $4.07022 \times 10^{14}$& $5.67746 \times 10^{14}$& $4.0706 \times 10^{14}$\\ [1ex]
\hline
Cs 6& $5.40694 \times 10^{14}$& $4.04946 \times 10^{14}$& $5.40651 \times 10^{14}$& $4.04955 \times 10^{14}$& $5.40633 \times 10^{14}$& $4.04989 \times 10^{14}$\\ [1ex]
\hline
\end{tabular}
\caption{In $g/cm^{3}$, the estimated expected values of the core and surface densities.}
\label{table:2}
\end{table}

We deduced that, for all the six compact stars under three different viable models, $\frac{d\rho}{{dr}}<0$, $\frac{dp_r}{{dr}}<0$ and $\frac{dp_t}{{dr}}<0$. For $r\rightarrow0$, we get
$$\frac{d\rho}{{dr}}\rightarrow0$$
$$\frac{dp_r}{{dr}}\rightarrow0$$
This is to be expected since these are diminishing functions, and we have a maximum density for small $r$ e.g. star core density $\rho_c=\rho(0)$.

\subsection{Validity of Different Energy conditions}
The Raychaudhuri equation for expansion \cite{ref45} has been used to obtain the general structure of the energy conditions. The essence of gravity is appealing with positive energy density, which cannot flow faster than the speed of light, as can be deduced from these conditions.
In Refs \cite{ref46,ref47,ref48,ref49,ref50,ref51}, the terms "Null-Energy conditions" $(NEC)$ and "Strong-Energy conditions" $(SEC)$ are thoroughly discussed.\\
\begin{itemize}
\item WEC $:\Rrightarrow$ $\rho \ge 0$, ${\rho + p_r} \ge 0$, ${\rho + p_t} \ge 0$,
\item NEC $:\Rrightarrow$ ${\rho + p_r} \ge 0$, ${\rho + p_t} \ge 0$,
\item SEC $:\Rrightarrow$ ${\rho + p_r} \ge 0$, ${\rho + p_t} \ge 0$, ${\rho + p_r + 2p_t} \ge 0$,
\item DEC $:\Rrightarrow$ $\rho \ge |{p_r}|$, $\rho \ge |{p_t}|$.
\end{itemize}
All these energy conditions for the compact stars are fulfilled for our feasible models for all the six odd star candidates.

\subsection{Tolman-Oppenheimer-Volkoff (TOV) Equation}
\begin{figure}[h!]
\centering
\includegraphics[width=40pc]{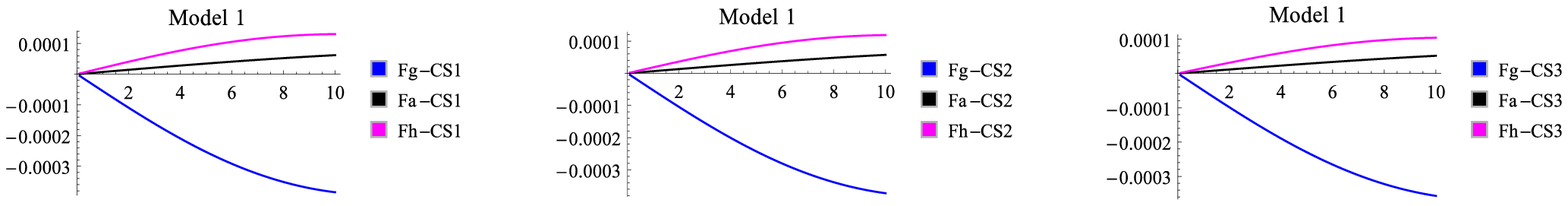}
\includegraphics[width=40pc]{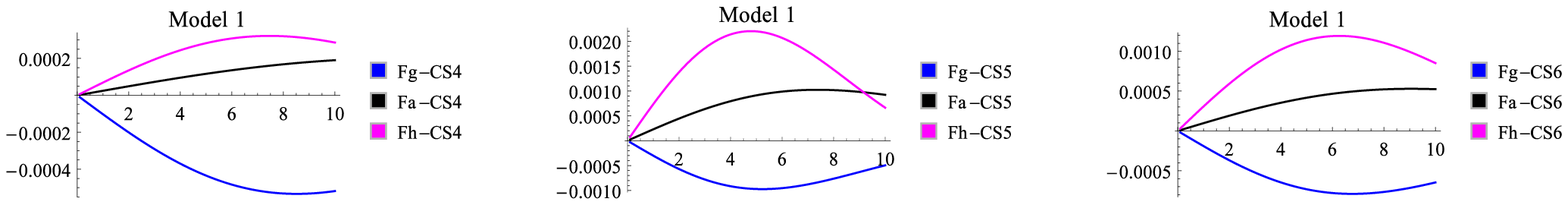}
\caption{The variation of gravitational force ($F_g$), hydrostatic force ($F_h$) and anisotropic force $(F_a)$ with respect to
the radial coordinate r (km).}\label{eqb}
\end{figure}

The generalization of TOV equation can be written as
\begin{equation}
\frac{{d{p_r}}}{{dr}} + \frac{{a'(p_r +\rho )}}{2} - \frac{{2({p_t} - {p_r})}}{r} = 0
\end{equation}
It can also be expressed in terms of magnetic, hydrostatic, and anisotropic forces
\begin{equation}
F_h + F_g + F_a= 0,
\end{equation}
which yields
$$F_g=- r(\rho+p_r)B , F_h=-\frac{{d{p_r}}}{{dr}}, F_a= -\frac{{2({p_t} - {p_r})}}{r}$$
We plot three different compact stars using these concepts, as seen in fig. (\ref{eqb})

Concerning the radial coordinate $r (km)$, we can see the difference of gravitational force ($F g$), hydrostatic force ($F h$), and anisotropic force $(F a)$. Model 1 is represented by the left plot, Model 2 by the middle plot, and Model 3 by the right plot.
\subsection{The Stability-Scenario}
Here, we look at the stability of compact star models under viable models in the $f(R,G)$ theory. To test our model's stability, we measure the radial and transverse speeds as follows,
$$\frac{{d{p_r}}}{{d\rho }} = v_{sr}^2$$
and
$$\frac{{d{p_t}}}{{d\rho }} = v_{st}^2$$
For stability, the radial as well as the transverse speed should obey the condition $v_{sr}^2\in[0,1]$ and $v_{st}^2\in[0,1]$.

\begin{figure}[h!]
\centering
\includegraphics[width=20pc]{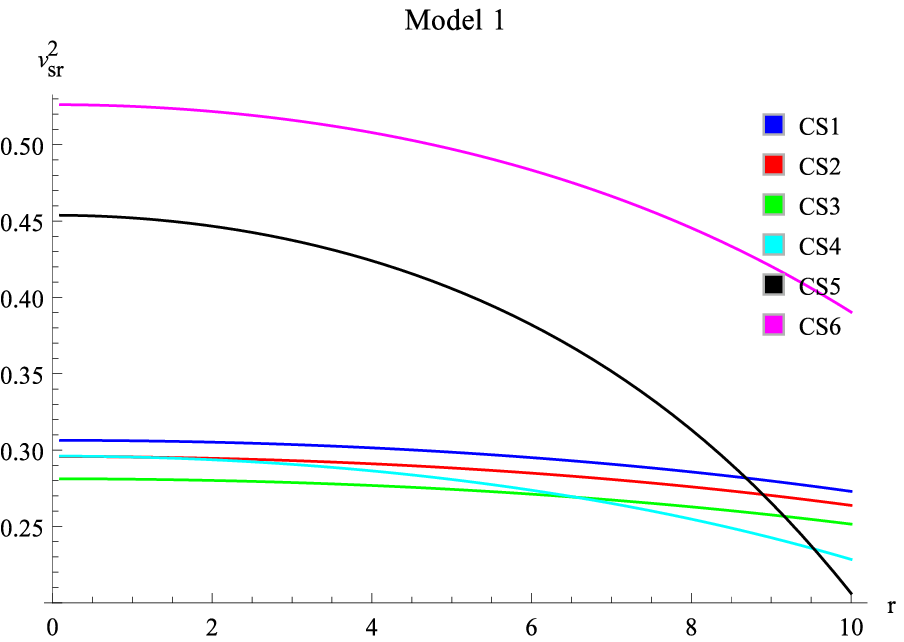}
\includegraphics[width=20pc]{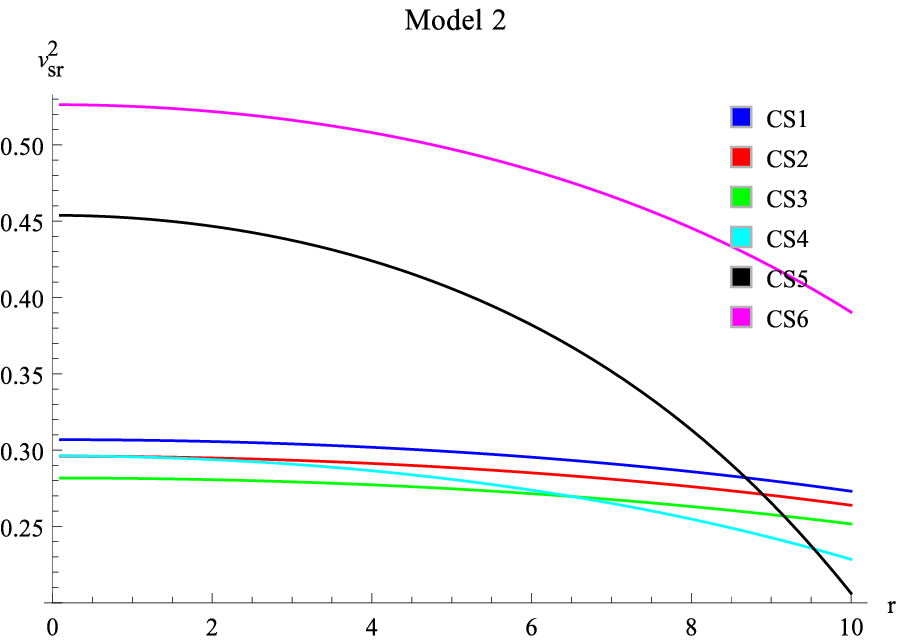}
\includegraphics[width=20pc]{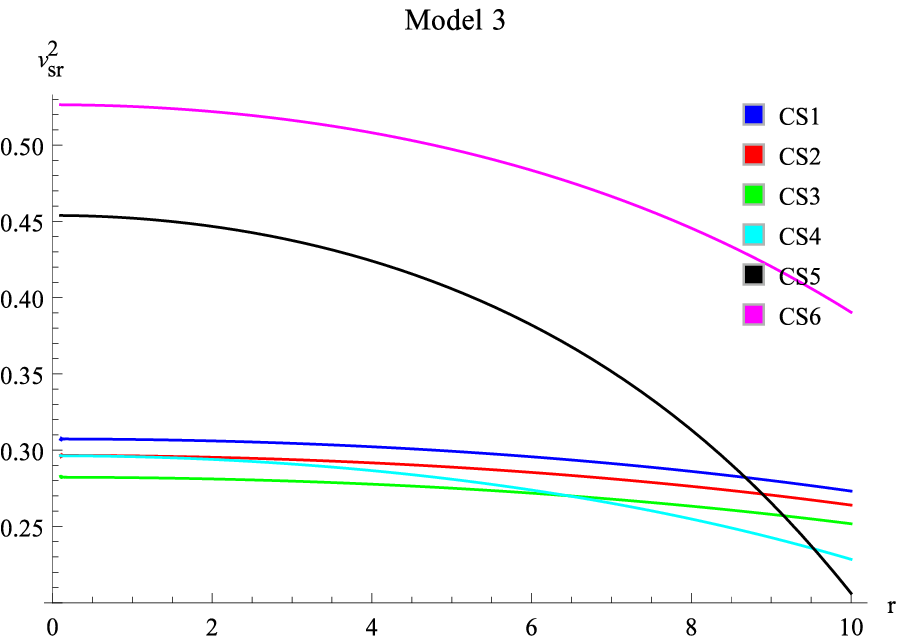}
\caption{Variations of $v_{sr}^2$ with respect radius $r$ (km) of the strange stars}\label{vsr}
\end{figure}

\begin{figure}[h!]
\centering
\includegraphics[width=20pc]{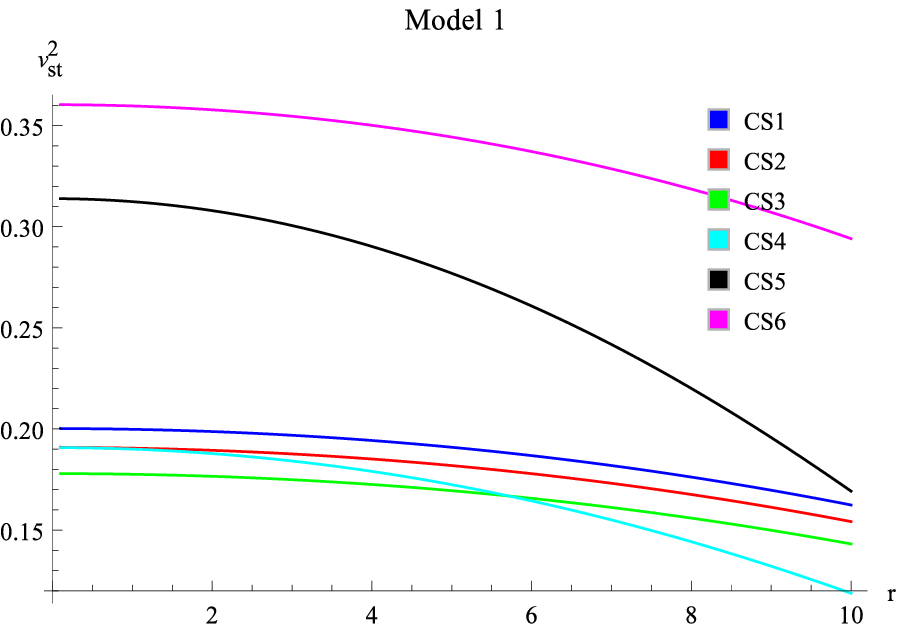}
\includegraphics[width=20pc]{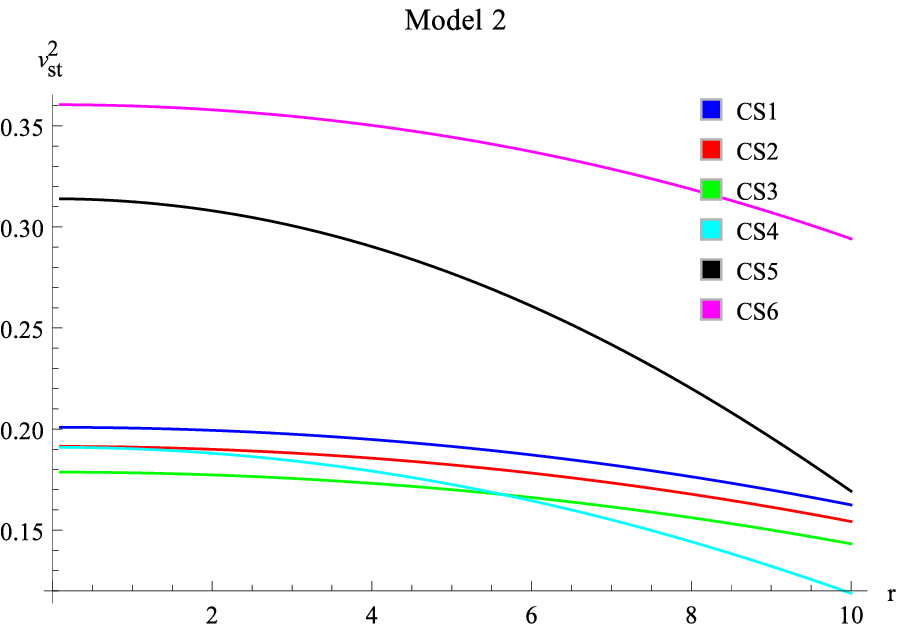}
\includegraphics[width=20pc]{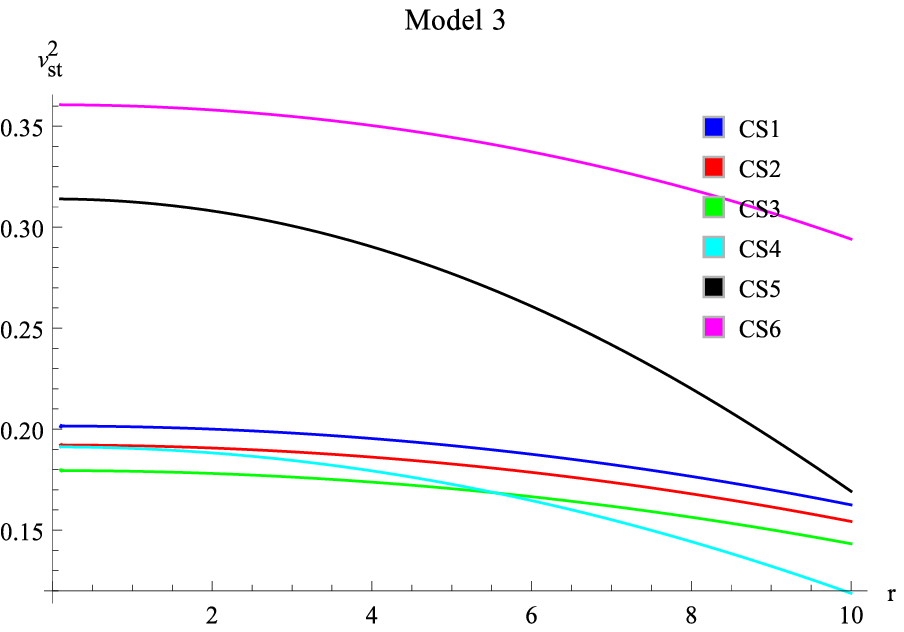}
\caption{Variations of $v_{st}^2$ with respect radius $r$ (km) of the strange stars}\label{vst}
\end{figure}

It is evident from Figure (\ref{vsr}) and (\ref{vst}), that the variations in radial and transversal sound speeds, for all the six types of odd star candidates, falls beyond the discussed stability bounds.

Similarly, we reached at
$$0< |v_{st}^2-v_{sr}^2|<1$$
so the stability is attained for compact stars in the $f(R,G)$ gravity models.
\subsection{Equation of State Parameters}
\begin{figure}[h!]
\centering
\includegraphics[width=20pc]{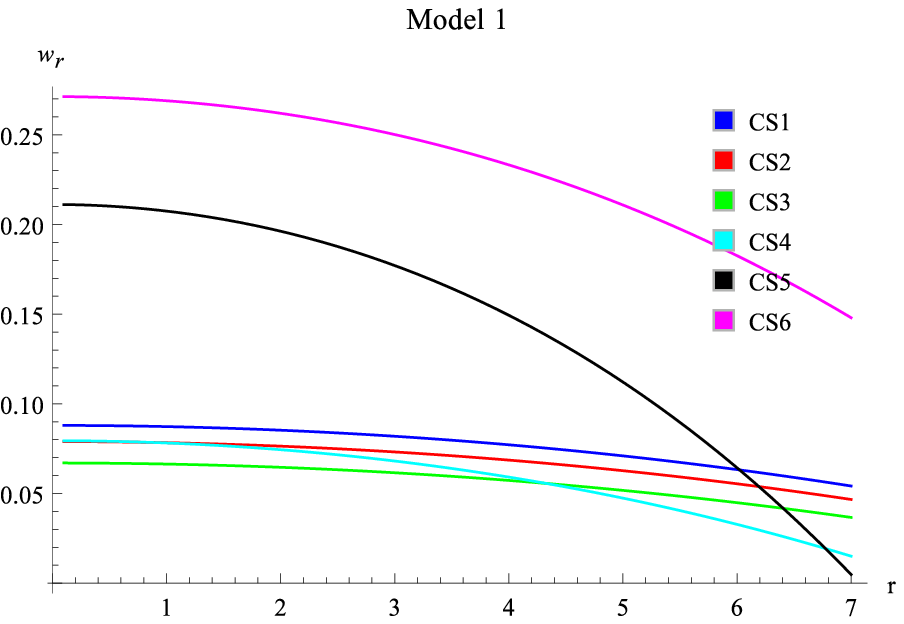}
\includegraphics[width=20pc]{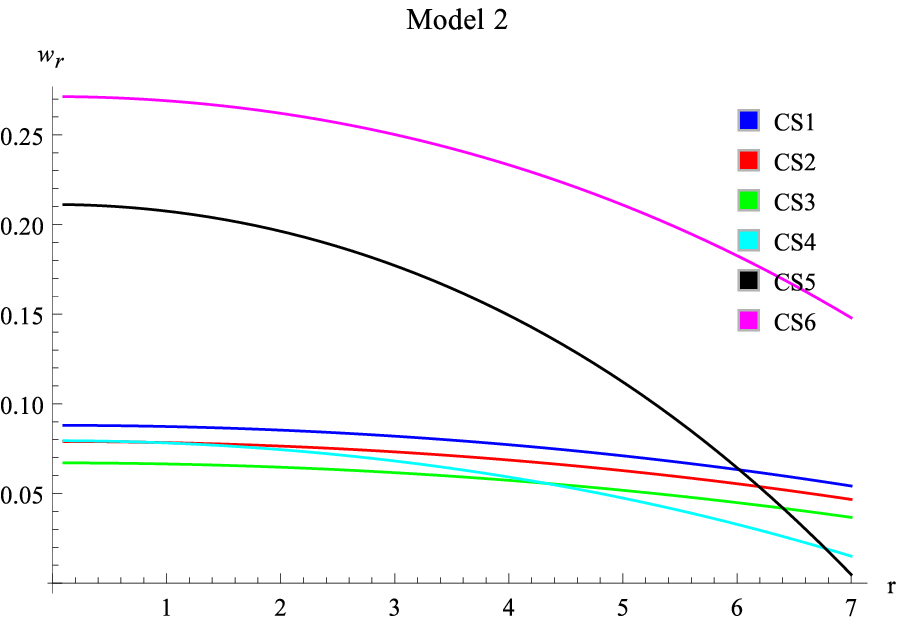}
\includegraphics[width=20pc]{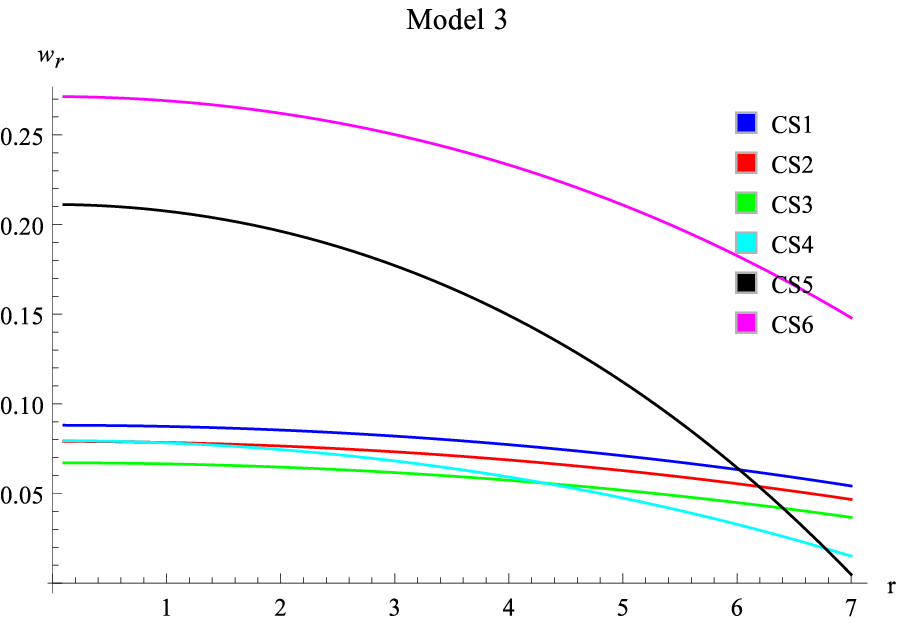}
\caption{The variations of the radial EoS parameter against the radial coordinate, $r$ $(km)$.}\label{wr}
\end{figure}

\begin{figure}[h!]
\centering
\includegraphics[width=20pc]{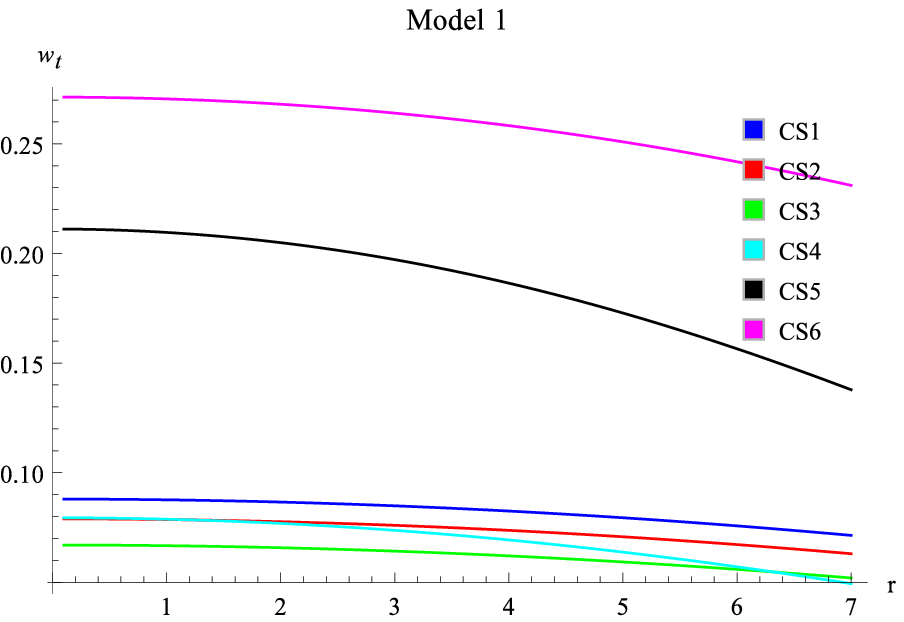}
\includegraphics[width=20pc]{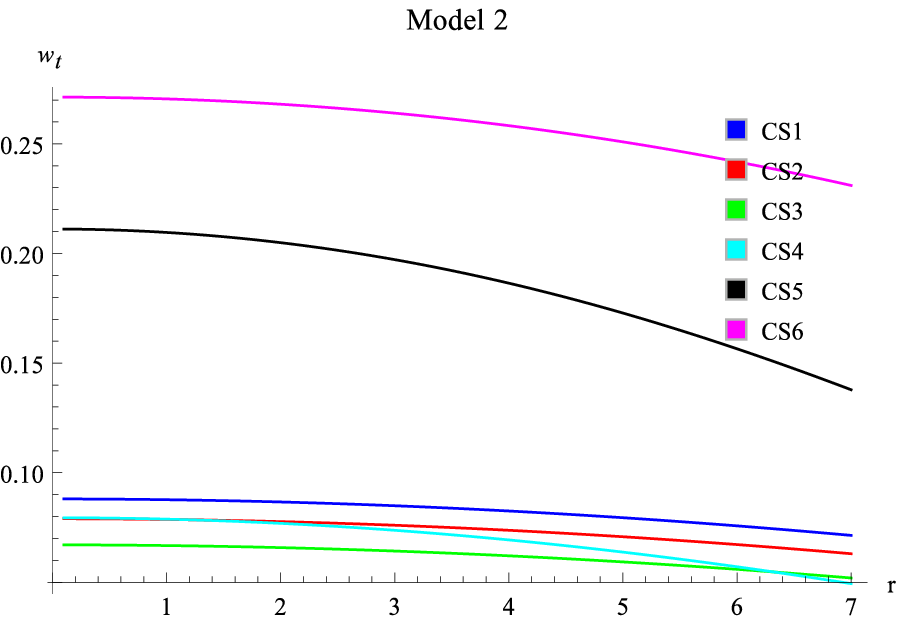}
\includegraphics[width=20pc]{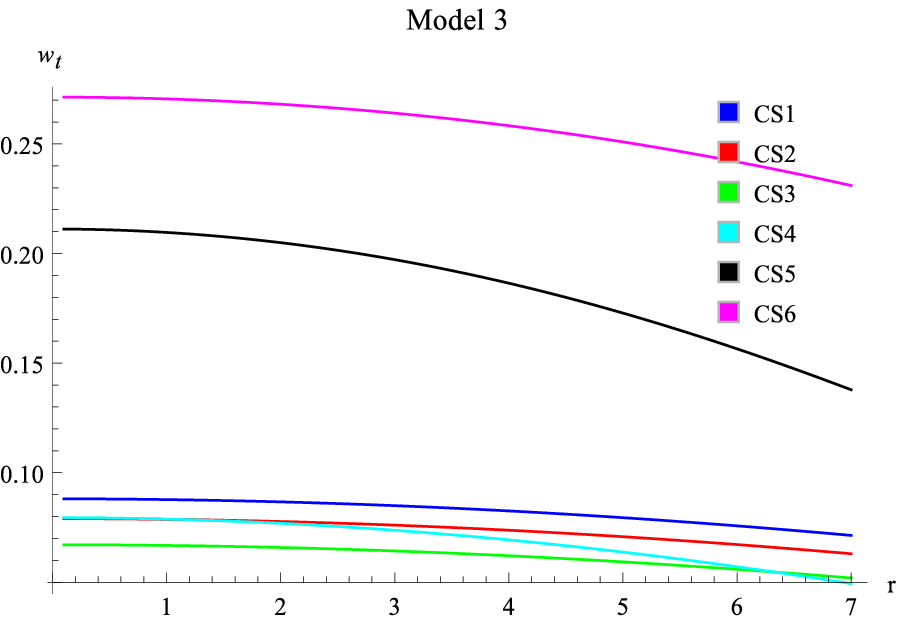}
\caption{The variations of the transverse EoS parameter against the radial coordinate, $r$ $(km)$.}\label{wt}
\end{figure}

Now for the anisotropic case, the equation of state $(EoS)$ are written as
$$p_r=w_r \rho$$
and
$$p_t=w_t \rho.$$
For which the limits are like $0<w_r<1$ and $0<w_t<1$. The behavior of $w_r$ and $w_t$ are shown graphically in fig. (\ref{wr}) and (\ref{wt}) respectively.

\subsection{The Measurement of Anisotropy}
\begin{figure}[h!]
\centering
\includegraphics[width=20pc]{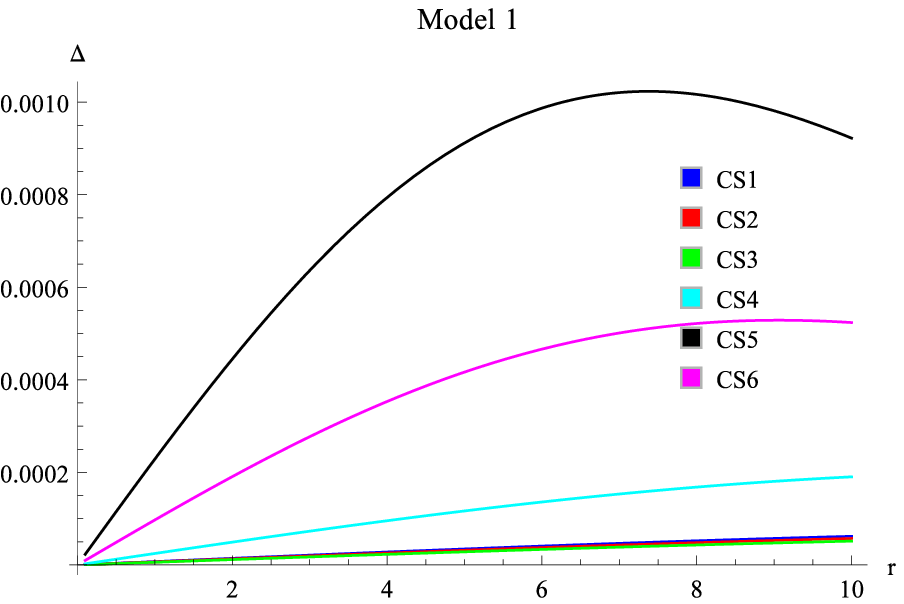}
\includegraphics[width=20pc]{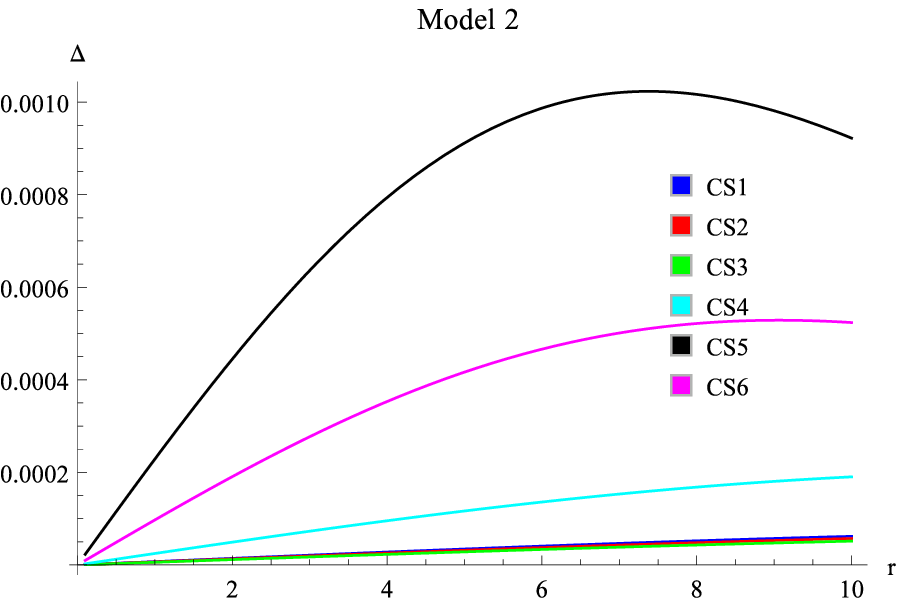}
\includegraphics[width=20pc]{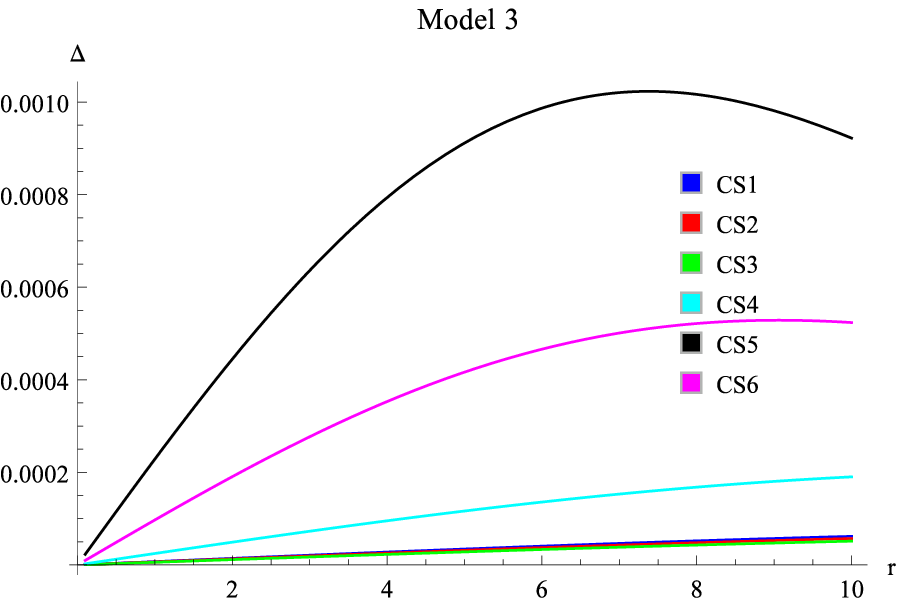}
\caption{the variations of anisotropic measure $\Delta$ against the radial coordinate, $r$ $(km)$.}\label{ani}
\end{figure}

The anisotropy is defined as
\begin{equation}
\Delta = \frac{2}{r}({p_t} - {p_r})
\end{equation}
We get $\Delta > 0$ while plotting the anisotropy, e.g. $p_t > p_r$, which indicates that the metric of anisotropy is oriented outward. Figure(\ref{ani}) illustrates these plots.

\section{Summary}
\vspace{0.5cm}
In this paper, we address the interior solution of compact stars by assuming that their internal structure is anisotropic in the modified gravity theory $f(R,G)$. In this respect, six compact stars, namely Her Stars $Her X-1$, $SAXJ1808.4-3658$, $4U1820-30$, $PSR J 1614 2230$, $VELA X-1$ and $Cen X-3$ were considered to explore their physical features in the $f(R,G)$ theory of gravity. The arbitrary unknown constants $A$, $B$, and $C$ were extracted from the smooth matching condition of the Schwarzchild exterior-metric to the interior-metric's analytic solutions. The matching condition enabled us to express the masses and radii of compact stars in terms of arbitrary constants and thus lead us to understand the nature and existence of these compact bodies. Based on their physical properties, the following conclusions about the anisotropic compact stars in the $f(R,G)$ gravity were drawn.
Compact stars have the equation of state parameters that are the same as an ordinary-matter distribution in the $f(R,G)$ gravity, highlighting that they are made up of ordinary matter. The matter/energy density, as well as radial and tangential pressure, were plotted as a function of radial coordinate $r$ which revealed that for all the six strange star candidates the density reaches its maximum limit when $r$ goes to zero. It confirms that the compact star's matter components are positive and finite in their interiors leading us to conclude that none of the six compact stars in this analysis have a singularity which further supports the theory that the cores of compact stars are extremely compact. The obtained results also verify as discussed in a recent proposed theory in Ref. \cite{ref52}.
It's also worth noting that for $p_{t}< p_{r}$, the direction of anisotropic force is inward implying $\triangle<0$, while in the reverse scenario i.e., $p_{t}> p_{r}$ the anisotropy become positive suggesting that the anisotropic force is being outward-directed. For six distinct compact stars, the graphical representation of $\triangle>0$ is presented in fig $14$.

For each of the six compact stars under investigation, the TOV and energy conditions are examined as revealed in Fig. \ref{vsr} and \ref{vst} respectively. Additionally, the inequality $|v^2_{sr}- v^2_{st}| < 1$ holds for all the six candidates indicating that the models under consideration seems to be potentially stable in $f(R,G)$ gravity.

\bibliographystyle{unsrt}
\bibliography{mybibcompactstars}
\end{document}